\documentclass[twocolumn, showpacs, amsmath,amsfonts,prb]{revtex4-1}
\usepackage{graphicx}
\usepackage{amsmath}
\usepackage{amssymb}
\usepackage{bm}
\usepackage{times}

\begin{document}
\title{Effect of photonic crystal stop-band on photoluminescence of \textit{a}-Si$_{1-x}$C$_x$:H}
\author{Mikhail~V.~Rybin${}^{1,2}$}
\email{m.rybin@mail.ioffe.ru}
\author{Alexander V. Zherzdev${}^{1}$}
\author{Nikolay A. Feoktistov${}^{1}$}
\author{Alexander B. Pevtsov${}^{1}$}

\affiliation{$^1$ Ioffe Institute, St.~Petersburg 194021, Russia\\
$^2$Department of Photonics of Dielectrics and Semiconductors, ITMO University, St.~Petersburg 197101,
Russia }

%\pacs{42.25.Fx, 42.70.Gi, 78.66.-w, 78.67.Pt \textit{\textbf{{\red{\textbf{check}}}} }}
\begin{abstract}
Effects associated with the change in the local density of photonic states in a periodic structure based on alternating \textit{a}-Si$_{1-x}$C$_x$:H and \textit{a}-SiO$_2$  amorphous layers forming a one-dimensional (1D) photonic crystal have been analyzed. 
The use of \textit{a}-Si$_{1-x}$C$_x$:H as the emitting material made it possible to examine the transformation of the photoluminescence spectrum contour that is comparable in width with the photonic stop-band. It was experimentally demonstrated that the emission is enhanced and suppressed in the vicinity of the stop-band. The relative intensities of the luminescence peaks at different edges of the stop-band vary with the detuning of the stop-band position and photoluminescence peak of a single \textit{a}-Si$_{1-x}$C$_x$:H film. The Purcell effect in the system under consideration was theoretically described by the method in which the local density of photonic states is calculated in terms of a 1D model. 
%A good agreement was obtained between the experimentally measured and calculated spectra. 
It was shown that the specific part of local density of states substantially increases at the long-wavelength (low-frequency) edge of the stop-band of a 1D photonic crystal as a result of the predominant localization of the electric field of the light wave in the spatial regions of \textit{a}-Si$_{1-x}$C$_x$:H which have a higher relative permittivity as compared with \textit{a}-SiO$_2$. 
\end{abstract}

\date{\today}
\maketitle

\section{Introduction}

The control over the spontaneous emission via modification of the local density of photonic states (photonic LDOS) due to the change in the spatial environment of an emitting center, i.e., to the Purcell effect \cite{purcell1946spontaneous}, has been extensively studied both theoretically and experimentally because of having a considerable application potential \cite{weisbuch2000overview,noda2007spontaneous}. One of widely used ways to modify the LDOS is to place the emitting center in the microcavities \cite{
kavokin2007microcavities,sauvan2013theory,rybin2016purcell,ding2016multidimensional}. Among other objects strongly modifying the LDOS should be mentioned the periodic structure of photonic crystals (PhCs), which can dramatically change the density of states to nearly zero within the photonic band gap (in the case of 3D PhCs) or to rather large values at stop-band edges \cite{yablonovitch1987inhibited, g101}. An important specific feature of PhC is the spatial extension of Bragg resonances, which enables modification of the emission rate from a material occupying a substantial volume in the structure \cite{benisty2012photonic,vos2015localization}. The increase in the density of states at the band edge makes it possible to observe, to name a few, such effects as the enhancement of the spontaneous emission \cite{tocci1996measurement,kuroda2009doubly,kuroda2010enhanced}, generation of harmonics \cite{scalora1997pulsed}, lasing \cite{nakamura1973optically, kopp1998low,kok2008lasing}, enhancement of luminescence under two-photon pumping \cite{markowicz2002enhancement}, increase in the incident photon-to-current conversion efficiency \cite{halaoui2005increasing}, enhanced Raman scattering \cite{inoue2008dramatic}, and control over exciton emission \cite{wu2014control}.

An important class of photonic crystals is constituted by planar layered structures (1D PhCs) or distributed Bragg reflectors. As a rule, the most common fundamental properties of PhCs can be understood for the example of 1D structures. Compared with 3D photonic crystals, the well-developed growth technologies of thin films can produce perfect planar periodic structures with prescribed precision. And, last but not least, layered structures can be comparatively large in size, which markedly facilitates experimental studies.

A theoretical analysis of the modification of the spontaneous emission rate is based on Fermi's golden rule and calculation of the photonic LDOS either via the eigenmodes of the PhC \cite{bendickson1996analytic,sprik1996optical}, or by calculation of the imaginary part of Green's function \cite{poddubny2012microscopic, yeganegi2014local}. Also used in the literature is the `indirect' method based on the Kirchhoff law for the equilibrium emission \cite{cornelius1999modification,bisson2011indirect}. This approach has been used for already more than 60 years to describe the luminescence in semiconductors, beginning from the pioneering communication by van Roosbroeck and Shockley \cite{van1954photon}.

The theory predicts the presence of two peaks in the PhC emission spectra, which correspond to different edges of a stop-band.
We notice that experimental data obtained in the case of radiative centers with a luminescence contour much wider than the stop-band width \cite{kuroda2010enhanced} demonstrates that 
the intensity of the short-wavelength peak differs from that of the long-wavelength one.
In other studies, spectra of centers with an emission line substantially narrower than the stop-band were examined \cite{
fogel1998spontaneous, medvedev2005enhancement}. To the best of our knowledge, no observations of a simultaneous change in the intensity of the peaks at the short- and long-wavelength edges of the stop-band due to the spectral detuning of the emitter wavelength and the stop-band (at comparable widths of the stop-band and the emitter) have been reported to date.

In this paper we present an experimental and theoretical analysis of the interaction conditions of radiative centers with the electromagnetic modes of a 1D photonic-crystal structure in the case in which the peak of the luminescence contour of emitting centers is tuned to approximately the center of the photonic stop-band (PSB) and the emission spectrum is comparable in width with the PSB. Multilayer structures constituted by alternating quarter-wave \textit{a}-Si$_{1-x}$C$_x$:H and \textit{a}-SiO$_2$ amorphous layers were chosen as objects of our study. An advantage of these materials is that high-contrast 1D PhCs possessing a comparatively wide PSB in the visible to the near-IR spectral range can be formed from the minimum number of layers (as small as 3-4 pairs). The role of radiative emitters is played in this structure by \textit{a}-Si$_{1-x}$C$_x$:H layers exhibiting a high-intensity photoluminescence (PL) in the visible spectral range at room temperature. The spectral positions of both the PL contour and the PSB can be easily widely varied by changing the content of carbon in the \textit{a}-Si$_{1-x}$C$_x$:H film. In addition, the full width at half-maximum of the PL contour of \textit{a}-Si$_{1-x}$C$_x$:H reaches a value of 0.6-0.7 eV, which is comparable with the PSB size for a 1D PhC. This opens up the opportunity of detecting emission peaks in the spectral ranges of both the long- and short-wavelength edges of the PSB in a single measurement and comparing these peaks. The theoretical analysis of the experimental data is based on using the method of Green's functions for calculating the LDOS and the Purcell factor in a 1D model.

The paper is organized as follows. Section II describes the fabrication technique and characterizes the samples under study. Section III contains experimental results. The 1D model used to calculate PL spectra is presented in Section IV. The general discussion is carried out in Section V.

%\section{Sample fabrication}
\section{Samples: fabrication and characterization}
%\section{Samples}

\begin{figure}[b]
\includegraphics{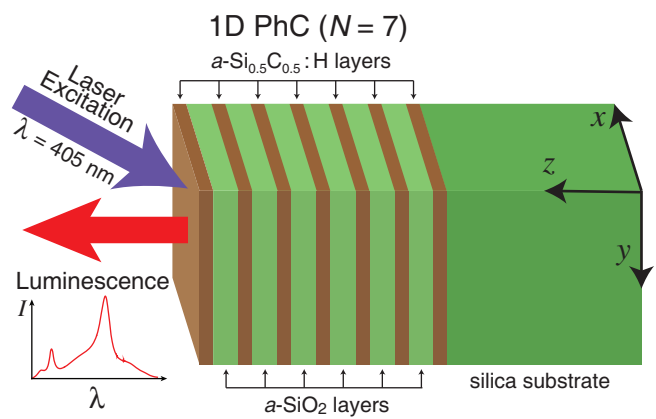}
\caption{ (color online) Schematic view of the 1D PhC structure composed of $N=7$ emitting
a-Si$_{0.5}$C$_{0.5}$:H layers and 6 a-SiO$_2$ layers on the silica substrate. The silica substrate is colored by dark green. The fabricated layers are colored by brown (a-Si$_{0.5}$C$_{0.5}$:H) and light green (a-SiO$_2$). The structure is excited by an oblique laser beam at $\lambda=405$ nm. The photoluminescence spectra are detected in normal direction.} \label{fig:Schematic}
\end{figure}

In our study, we examined 1D PhCs schematically shown in Fig. ~\ref{fig:Schematic}. The samples have the form of alternating \textit{a}-Si$_{1-x}$C$_x$:H and \textit{a}-SiO$_2$ layers deposited on commercially available silica substrates. The structures were fabricated by the plasma-enhanced chemical vapor deposition (PECVD) technique in a capacitive reactor (electrode diameter $\sim$8 cm) at various relative flows of silane (SiH$_4$), methane (CH$_4$), and oxygen (O$_2$).

%%%%%%%%%%%%%%%%%%%%%%%%%%%%
%% Figure 1
%
%%%%%%%%%%%%%%%%%%%%%%%%%%%%

%%%%%%%%%%%%%%%%%%%%%%%%%%%%
%% Figure 2
\begin{figure*}[!t]
\includegraphics{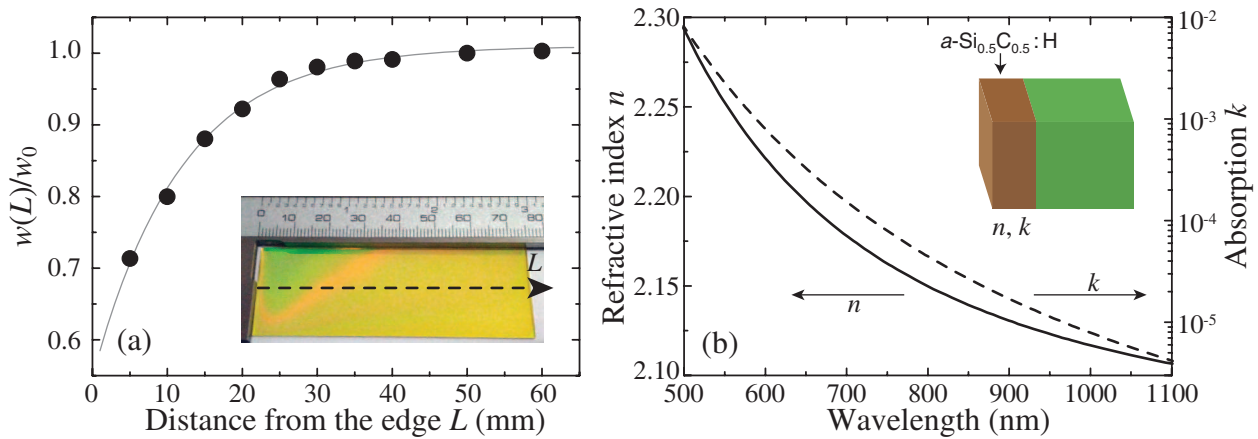}
\caption{(color online) Sample characterization. (a) Relative layer thickness $w/w_0$ vs. the distance form the sample edge, $L$ ($w_0$  is the thickness of the growing film at the center of the PE CVD reactor). The gray line is a guide for eyes only. A photograph of the sample is show in the insert. (b) The dispersion of the refractive index $n$ (solid curve) and absorption coefficient $k$  (dashed curve) for a-Si$_{0.5}$C$_{0.5}$:H determined from experimentally measured ellipsometric data. The layer thickness is $w=400$ nm. } \label{fig:Sample}
\end{figure*}
%
%%%%%%%%%%%%%%%%%%%%%%

All the structures were constituted by seven quarter-wave \textit{a}-Si$_{1-x}$C$_x$:H layers and six quarter-wave \textit{a}-SiO$_2$ layers and were fabricated in a single technological cycle without exposure of the samples to air between separate procedures. To deposit \textit{a}-SiO$_2$, silane was strongly diluted with oxygen. Other details of the technological process have been described in our previous communications \cite{dukin2003polarization,dukin2006polarization,dukin2008eigenmode, medvedev2014planar}. The optical thickness of the layers in the structure were monitored in situ by the interference pattern appearing when the intensity of the laser beam reflected from the surface of a growing film was detected. Fused silica plates $20\times 75$ mm were served as substrates. The specificity of the samples consists in that they have a noticeable (approximately the same for all the layers) thickness gradient along the long side L of the substrate (Fig .~\ref{fig:Sample}a) sufficient for spectrally shifting the PSB center of the structure within the range on the order of 100 nm. The appearance of the gradient is due to the large longitudinal size (compared with the diameter of the electrodes in the PECVD reactor) of the samples, which resulted in that a considerable part of the substrate was situated in the region of a nonuniform burning of the glow discharge, which, in turn, led to a change in the growth rate of separate parts of the films away from the center of the electrodes.

The dispersion of dielectric index was determined by the method of spectral ellipsometry on an M-2000 instrument manufactured by J.A. Woollam Co., Inc.
Measurements were made on single \textit{a}-Si$_{1-x}$C$_x$:H and \textit{a}-SiO$_2$ films with the thickness of 400 nm. The dependences obtained for the ellipsometric angles were used to calculate the spectral dependences of the real $n$ and imaginary $k$ parts of the refractive index in terms of the model that considers air, homogeneous isotropic film, and semi-infinite quartz substrate. The results of calculation for the \textit{a}-Si$_{0.5}$C$_{0.5}$:H film are shown in Fig.~\ref{fig:Sample}b. The plot for $n$ is well described by the Cauchy's dispersion formula $n=2.058+58817/\lambda^2$. The value of absorption $k$ in the spectral range 500-1100~nm that is important for the study is relatively weak.
The optical constants $n$ and $k$ for \textit{a}-SiO$_2$ films obtained under our technological conditions are in agreement with the well-known literature data \cite{malitson1965interspecimen} for fused silica and are characterized by the negligible absorption and dispersion  in the spectral range under study.
%The optical constants $n$ and $k$ for \textit{a}-SiO$_2$ films obtained under our technological conditions are in agreement with the well-known literature data \cite{malitson1965interspecimen} for fused silica and are characterized by the absence of absorption and a dispersion that is negligible in the spectral range under study.
Therefore, we used a constant value  $n_{\mathit{a}\mathrm{-SiO}_2}=1.46$ in our calculations.

\section{Experimental}

The transmission and luminescence spectra of the films and structures were measured with an Ocean Optics USB4000 mini-spectrometer in the spectral range 500-1100~nm. To avoid the appearance of the anti-Stokes PL wing, the emission was excited with a continuous-wave semiconductor laser at a wavelength of 405 nm, (the corresponding phonon energy exceeds the electronic energy gap width of \textit{a}-Si$_{1-x}$C$_x$:H at any value of $x$). In addition, according to published data this wavelength falls outside the excitation spectrum of the PL associated with oxygen vacancies in \textit{a}-SiO$_2$  films \cite{sakurai1999characteristic}. The power density of laser light did not exceed 40~mW/mm$^2$. The PL spectra were corrected with consideration for the spectral sensitivity of the equipment. To suppress the Fabry-Perot interference on the thickness of a single film, the back surface of the quartz substrate was ground. It is also noteworthy that the PL signal was not detected for single \textit{a}-SiO$_2$ films obtained in the chosen PECVD technological modes, within the sensitivity limits of our measurement apparatus. No PL signal was detected, either, from the silica substrates we used.

\begin{figure}[t]
\includegraphics{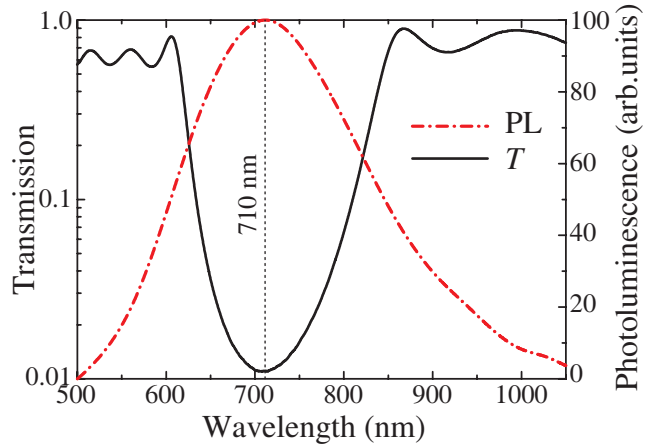}
\caption{ Transmission spectrum of \textit{a}-Si$_{0.5}$C$_{0.5}$:H/\textit{a}-SiO$_2$ one-dimensional PhC on the logarithmic scale (black solid curve) and the photoluminescence spectrum of a single \textit{a}-Si$_{0.5}$C$_{0.5}$:H film (red dash-and-dotted curve). Wavelength 710 nm is marked by vertical dotted line. } \label{fig:PL0andT710}
\end{figure}

It is known that, as the content of carbon ($0<x<1$) in \textit{a}-Si$_{1-x}$C$_x$:H films increases, the luminescence peak of this material gradually shifts to shorter wavelengths (to approximately 2.5 eV or 500 nm), and the refractive index $n$ decreases by more than a factor of 2, from $n = 4$ (for $x = 0$) to $n < 2$ (for $x > 0.6$) \cite{siebert1987photoluminescence,carius2000optical,giorgis2000optical}. In a set of technological experiments, we determined the ratio between the CH$_4$ and SiH$_4$ flows at which the maximum of the emission spectrum of a single \textit{a}-Si$_{1-x}$C$_x$:H reference film (synthesized under the same technological parameters as those at which a multilayer structure was fabricated) approximately coincided with the center of the PSB in a 1D PhC formed by alternating quarter-wave \textit{a}-Si$_{1-x}$C$_x$:H and \textit{a}-SiO$_2$ layers (Fig.~\ref{fig:PL0andT710}). 
Samples with a lower carbon content have the PL band narrower than the stop-band and possess a small quantum yield at room temperature, hence they were unsuitable for our task. Samples with a higher carbon content have a weak dielectric contrast (a ratio of the neighbor layer dielectric indices). In this case, in order to synthesize the high quality 1D PhCs with sharp edges of the stop band it is necessary to form a large number of periods, that is a technological problem.
At the same time, a 50\% content of carbon in \textit{a}-Si$_{1-x}$C$_x$:H film ($x = 0.5$)  corresponds to the refractive index of $n = 2.17$ and the peak in PL spectrum lying at a wavelength of 710 nm (Fig. ~\ref{fig:PL0andT710})\cite{siebert1987photoluminescence}. 
In this case, the width of the luminescence band exceeds the spectral size of the PSB, which enabled us to simultaneously detect in our experiments the emission intensity in the spectral ranges corresponding to the short- and long-wavelength PSB edges.

%%%%%%%%%%%%%%%%%%%%%%%%%%%
% Fig. 3
%
%%%%%%%%%%%%%%%%%%%%%%%%%%%

%%%%%%%%%%%%%%%%%%%%%%%%%%%%
%% Figure 4
\begin{figure}[!t]
\includegraphics{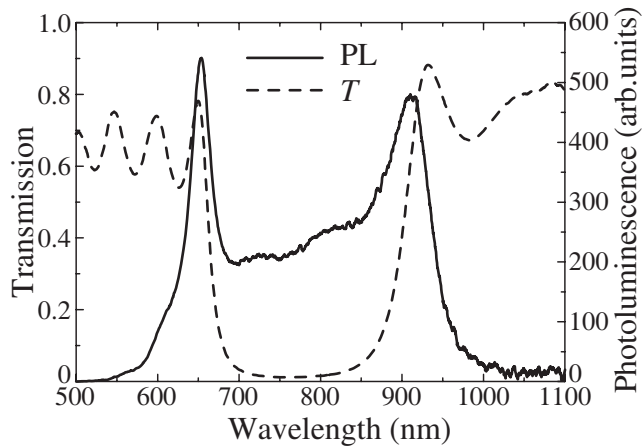}
\caption{Transmission (dash curve) and photoluminescence (solid curve) spectra of
the \textit{a}-Si$_{0.5}$C$_{0.5}$:H/\textit{a}-SiO$_2$ one-dimension PhC with PSB centered at $\lambda_0=765$ nm. }
\label{fig:PLandT}
\end{figure}
%
%%%%%%%%%%%%%%%%%%%%%%

The procedure by which the transmission and PL of the PhC structures under study is measured consists in the following. Light from an incandescent lamp was focused onto a sample into a spot 1 mm$^2$ in size. 
Emitted light was collected by lens with focal length of 100 mm and the beam cross-section was limited with a small iris diaphragm. As a result, the numerical aperture of the optical system did not exceed 0.025.
A laser beam for PL excitation was focused into the same region. Additionally, the optical scheme of a confocal microscope was assembled in front of the input optical fiber of the spectrometer to select a fixed uniformly illuminated surface area ($100\times 100$ $\mu$m) by using a system of adjustable crossed slits. The scheme enabled measurements of the transmission and PL spectra from the same region of a 1D PhC sample under study.

Let us consider the main specific features of the spectra we obtained. As an example, we present the spectra of a PhC structure with a PSB centered at 765 nm (Fig.~\ref{fig:PLandT}). The transmission spectrum (dashed curve in Fig.~\ref{fig:PLandT}) shows an extended, about 120-140 nm wide, dip corresponding to the PSB of 1D PhCs. The oscillations outside the PSB band are due to the Fabry-Perot interference over the whole thickness of the structure under study. The PL spectrum (solid curve in Fig.~\ref{fig:PLandT}) demonstrates two peaks with different intensities, which lie near the short- and long-wavelength edges of the PSB. The emission intensity substantially decreases in the PSB range in the structures under study, but it still has a noticeable value, which is due to the finite number of PhC layers, additional contribution to the PL from the outer surface of the upper \textit{a}-Si$_{0.5}$C$_{0.5}$:H film (adjacent to air), and imperfection of the structure. Outside the PSB, the PL intensity rapidly decreases, with the band wings on both the short- and long-wavelength sides falling substantially more steeply, compared with the edges of the PL contour of a single \textit{a}-Si$_{0.5}$C$_{0.5}$:H film (Fig.~\ref{fig:PL0andT710}). Comparison of the transmission and PL spectra shows that the width of the PL peaks approximately correspond to the widths of the interference peaks in the transmission spectra. The PL peaks are spectrally shifted into the PSB relative to the interference peaks in the transmission spectrum, which delimit the PSB. We notice that this situation differs from that in planar microcavities (1D PhCs with a spatial defect) in which, under the conditions of weak coupling between the electromagnetic field and the emitting system, the contour profile of the spontaneous emission from the active layer of the microcavity in the range of the eigenmode frequency nearly coincides with the profile of the transmission contour of the microcavity structure \cite{dukin2000optical}.

%%%%%%%%%%%%%%%%%%%%%%%%%%%%
%% Figure 5
\begin{figure*}[!t]
\includegraphics{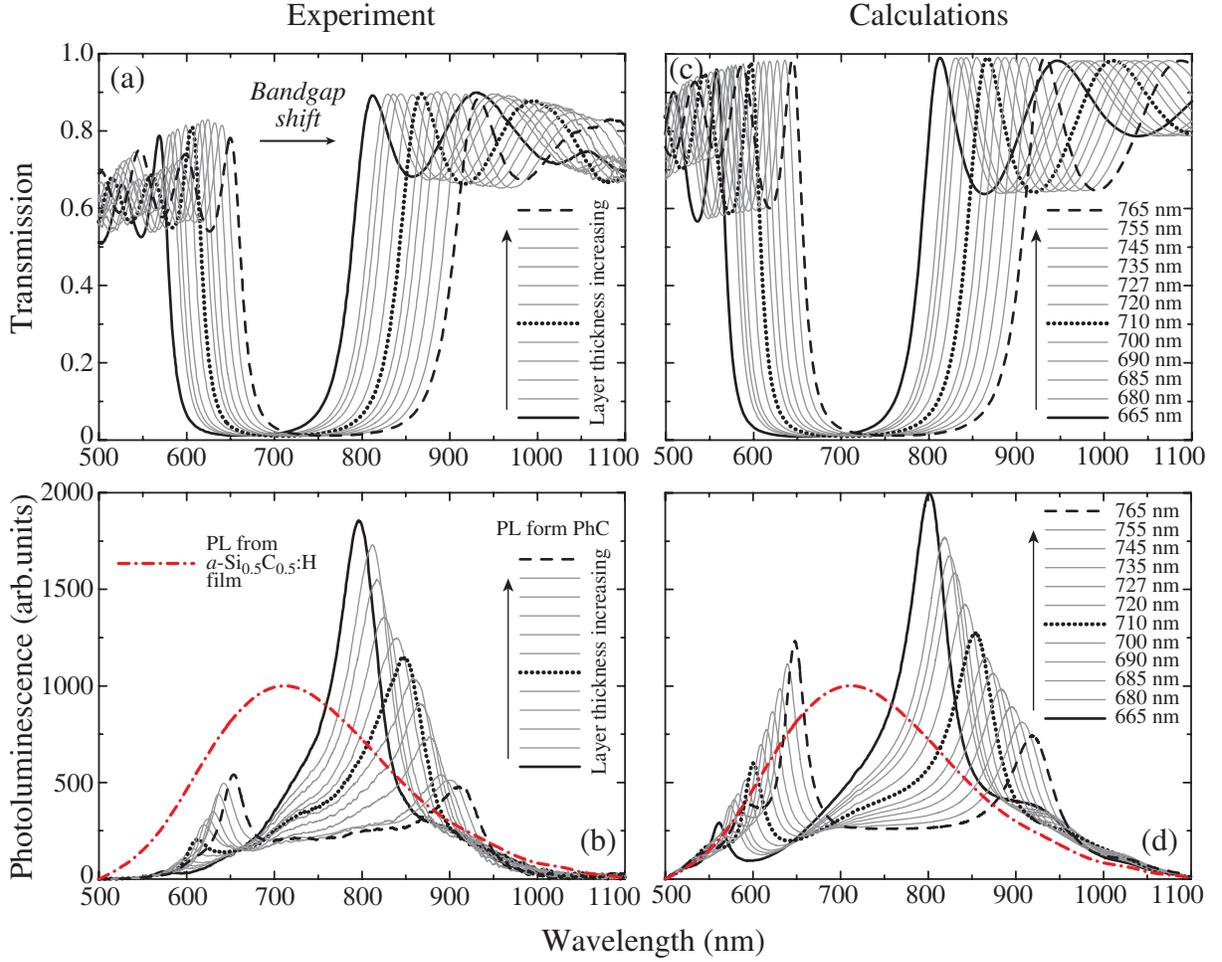}
\caption{(color online) Transmission and photoluminescence spectra of the \textit{a}-Si$_{0.5}$C$_{0.5}$:H/\textit{a}-SiO$_2$ one-dimension PhC as a function of the PSB center. Transmission spectra: experimentally measured (a) and calculated (c). PL spectra: experimentally measured (b) and
calculated (d). The red dash-and-doted curve is PL spectrum of {a}-Si$_{0.5}$C$_{0.5}$:H film 
with thickness of 400 nm.}
\label{fig:ExpCalc}
\end{figure*}
%
%%%%%%%%%%%%%%%%%%%%%%

Now we turn to the effects associated with the change in the thickness of the layers forming the 1D PhC under study. For this purpose, we successively made measurements from different spots of a sample, which correspond to various thicknesses of layers forming the structure (Fig. ~\ref{fig:Sample}a).
In particular, depending on location of the photonic stop band center $\lambda_0$, the thickness $w$ of separate quoter-wavelength layers of the structures were following: at $\lambda_0 =760$ nm, $w= 88.5$ nm for \textit{a}-Si$_{1-x}$C$_x$:H layer and $w =131$ nm for \textit{a}-SiO$_2$ layer, the total thickness of 13-layer structure is $1405$ nm; at $\lambda_0 =665$ nm, $w=76$ nm for  \textit{a}-Si$_{1-x}$C$_x$:H layer and $w =114$ nm for \textit{a}-SiO$_2$ the total thickness of 13-layer structure is $1215$ nm. The relative change in sample thickness $\Delta w/w_0$ (for $\lambda_0 = 765$ nm) is 13.5\%. 

In the course of a measurement, a sample was spatially shifted $\approx 1.5$ mm along direction $L$ (along the long side of the substrate, see Fig.~\ref{fig:Sample}a) to such a distance that the PSB center $\lambda_0$ shifted by approximately 10 nm (the PSB shift was monitored by directly measuring the transmission spectrum at a chosen spot). 
A set of transmission spectra measured as the position of the PSB center was varied is presented in Fig.~\ref{fig:ExpCalc}a. The black solid line corresponds to the shortest wavelength position of the PSB ($\lambda_0=665$ nm), the black dotted line corresponds to the PSB whose center $\lambda_0=710$ nm coincides with the PL peak from a single \textit{a}-Si$_{0.5}$C$_{0.5}$:H film (see Fig. ~\ref{fig:PL0andT710}), and the black dashed line corresponds to the longest wavelength PSB ($\lambda_0=765$ nm) under study. The intermediate spectra are represented by gray lines.

Let us now describe the behavior of the PL spectra measured from the same spots as the corresponding transmission spectra (Fig. ~\ref{fig:ExpCalc}b). For the region corresponding to $\lambda_0=665$ nm, a strongly asymmetric line is observed with high-intensity peak at $\lambda_2=795$ nm and a short-wavelength wing extended into the PSB region, with a weak peak present on this wing at $\lambda_1=572$ nm. As the PSB center $\lambda_0$ shifts to longer wavelengths, a noticeable red shift of the peaks at $\lambda_1$ and $\lambda_2$ is observed, with the intensity of the short-wavelength peak increasing and that of the long-wavelength peak, by contrast, decreasing. At a position of the PSB center $\lambda_0=755$ nm, the intensities of both peaks become equal. As $\lambda_0$ increases further, the intensity of the short-wavelength peak is found to be higher than that of the long-wavelength peak. It can be seen in Fig.~\ref{fig:ExpCalc}b that the transformation of the intensity of the PL peaks from the 1D PhC, observed as the PSB spectrally shifts, does not follow quantitatively the variation of the intensity of the PL spectrum of a single \textit{a}-Si$_{0.5}$C$_{0.5}$:H layer, represented by the red dash-dotted curve in Fig.~\ref{fig:ExpCalc}b.

\section{Theory}

\subsection{Transmission}

The light propagation in layered structures can be conveniently described in terms of the 1D model determined by the direction of the incident wave. In the present study, we use the transfer-matrix method \cite{g112}, which relates fields in different parts of the space. Here we assume that the wave vector is directed along the normal to the layer boundaries. Within a uniform layer, the electric field can be represented as the sum of two waves, $E(z)=E^{(+)}e^{ik_0 n z}+E^{(-)}e^{-ik_0 n z}$, where $E^{(+)}$, $E^{(-)}$ are the amplitudes of solutions in the form of plane waves propagating in the positive and negative directions of the  $\hat {z}$ axis,  $k_0=\omega/c$, $\omega$ is frequency, and $n$ is the complex refractive index in the layer. At any two points $z$ and $z_0$ within a layer, the field amplitudes are related by the matrix
\begin{equation}
\left(\begin{array}{c}
E^{(+)}(z)\\
E^{(-)}(z)
\end{array}\right)=\left(\begin{array}{cc}
e^{ik_0 n (z-z_{0})} & 0\\
0 & e^{ik_0 n(z_{0}-z)}
\end{array}\right)\left(\begin{array}{c}
E^{(+)}(z_{0})\\
E^{(-)}(z_{0})
\end{array}\right).
\label{eq:TrMatPro}
\end{equation}
In addition, Maxwell's boundary conditions relate the fields on both sides of the interface. This relationship can be written as the matrix
\begin{equation}
\left(\begin{array}{c}
E^{(+)}(z'+\delta)\\
E^{(-)}(z'+\delta)
\end{array}\right)=
\left(\begin{array}{cc}
\frac{n_{+}-n_{-}}{2n_{+}} & \frac{n_{+}+n_{-}}{2n_{+}}\\
\frac{n_{+}+n_{-}}{2n_{+}} & \frac{n_{+}-n_{-}}{2n_{+}}
\end{array}\right)
\left(\begin{array}{c}
E^{(+)}(z'-\delta)\\
E^{(-)}(z'-\delta)
\end{array}\right).
\end{equation}
Here, $\delta$ is a positive quantity that tends to zero, $n_\pm$ is the complex refractive index at $z\pm\delta$. As a result, a knowledge of the amplitudes  $E^{(+)}$ and $E^{(-)}$ at a point of space allows us to find the amplitudes at an arbitrary point of space via successive actions of transfer matrices. There also is a stronger statement: knowledge of any two amplitudes  $E^{(+)}$ and $E^{(-)}$ (with the exception of amplitudes of waves within the same layer if these waves propagate in the same direction) makes it possible to find amplitudes at any point of space. In particular, setting the unit amplitude of the field incident onto the structure from one side and assuming that the field incident from the other side is zero, we can find all field amplitudes, including the amplitude of transmission and reflection.

%%%%%%%%%%%%%%%%%%%%%%%%%%%%
%% Figure 6
\begin{figure}[!t]
\includegraphics{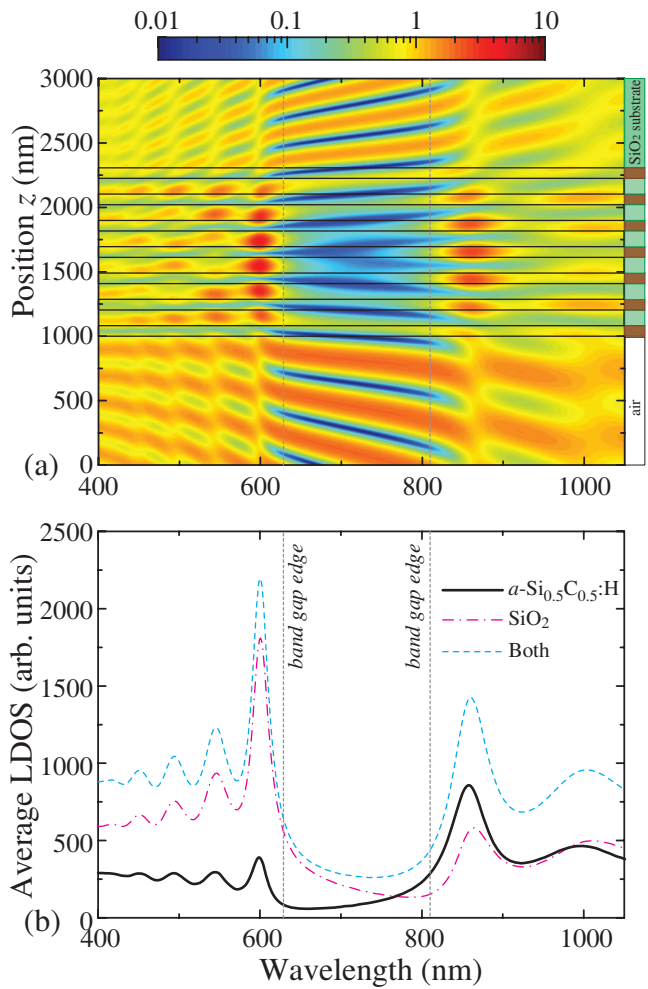}
\caption{(color online) One-dimensional Purcell factor distribution in the
\textit{a}-Si$_{0.5}$C$_{0.5}$:H/\textit{a}-SiO$_2$ PhC with $\lambda_0=710$ nm. (a) Calculated 1D Purcell factor as a function of the emitter position and its wavelength. The layer structure is shown in the right-hand part (air is white; \textit{a}-Si$_{0.5}$C$_{0.5}$:H is purple; and SiO$_2$ is magenta). The layer boundaries are
shown by the black horizontal lines. (b) The frequency dependence of the 1D photonic LDOS averaged over
\textit{a}-Si$_{0.5}$C$_{0.5}$:H layers (black solid curve), over \textit{a}-SiO$_2$ (magenta
dash-and-dotted curve), and over the whole structure (blue dashed curve). The vertical black dotted lines indicate the PBS edges.} \label{fig:LDOS}
\end{figure}
%
%%%%%%%%%%%%%%%%%%%%%%

When transmission spectra are calculated, it is necessary to determine whether the interface between the silica substrate and air should be taken into account. It results in that there appear Fabry-Perot oscillations with a period substantially shorter than 1~nm, which exceeds the resolution of the spectral instrument used in the study. In addition, oscillations of this kind are certainly broadened due to the beam divergence and in the simulation we averaged the spectra over an interval of 2 nm. These spectra differ from those calculated for a semi-infinite silica substrate only slightly. Therefore, we perform further calculations for the model with a semi-infinite silica substrate.

We use the transfer-matrix method to calculate transmission spectra for structures with quarter-wave layers for the wavelengths $\lambda_0$ in the range from 650 to 800~nm with a step of 1~nm. Figure~\ref{fig:ExpCalc}c shows a part of the spectra that correspond to the experimentally measured data presented in Fig. ~\ref{fig:ExpCalc}a. As a result, we determine the geometric parameters of the structures for which PL spectra is calculated in what follows.

\newpage

\subsection{Emission spectra}

Emission of light being inherently 3D process can be calculated by using a Green's tensor (i.e., a Green function in tensor form) of wave equation for electric field, and the imaginary part of its trace is known to define photonic LDOS \cite{novotny2012principles}. To the best of our knowledge, there is no analytical theory allowing to yield the photonic LDOS for an arbitrary structure, however for several types of photonic structures Green's tensor can be calculated in a form of infinite series. For example, it was succeeded for a stack of dielectric layers \cite{novotny1997allowed,wubs2002local} and array of cylindrical rods \cite{fussell2004three}. The main computational problem is that the Green's tensor is related to a spherical wave while the field in a layered structure or cylindrical rods is natural to expand in a plane wave or cylindrical wave basis, respectively. In the case of more complicated structures the Green's tensor can be calculated numerically, e.g., by using of finite-difference time-domain method \cite{hermann2002modified} or  by dehomogenization of media into coupled dipoles \cite{prosentsov2007local}. An alternative way is to consider a lower dimensional problem and solve it with appropriate 1D or 2D Green function in basis of cylindrical \cite{asatryan2001two} or plane waves \cite{yeganegi2014local}.

Because of the layered sample is quasi-1D object, we use the later method and assume that the 1D model is applicable to describe the features observed in the measured emission spectra.
According to Fermi's golden rule, it is necessary to also take into account, when calculating the decay rate $\gamma$ at which an emitting center relaxes from the excited (initial) to the ground (final) state with emission of a photon, the photonic LDOS which appears in the form of the Purcell factor $f$, i.e., $\gamma=f\,\gamma_0$. It is known that the LDOS can be expressed in terms of the imaginary part of Green's function, $\mathrm{Im}(G(z,z))$ \cite{economou2006green,novotny2012principles}. Applying the matrix approach\cite{rybin2016purcell}, we find for Green's function the following expression
\begin{equation}
G(z,z;k)=-\frac{i}{2k}\left(\frac{1+r_{-}+r_{+}+r_{+}r_{-}}{1-r_{+}r_{-}}\right),
\end{equation}
Here, $r_\pm$  is the complex reflectance at point $z$ for the wave propagating toward positive (or negative) values with respect to the $\hat{z}$ axis, which is calculated by the transfer-matrix method. Accordingly, the Purcell factor is calculated as the ratio between the photonic LDOS for a spot in the structure and the photonic LDOS for a spot in free space. We plotted the distribution of the 1D Purcell factor $f(\lambda, z)$ in relation to the wavelength $\lambda$ and the spatial position $z$ in the structure. This dependence is shown in Fig.~\ref{fig:LDOS}a for the structure with $\lambda_0=710$ nm. Neglecting the re-emission of light in the photonic crystal, we find the emission gain by integration within the whole structure
\begin{equation}
I(\lambda) = \alpha \int  f(\lambda, z) \gamma_0(\lambda, z) \mathrm{d}z,
\end{equation}
where $\alpha$ is the proportionality factor. If the emitting centers are uniformly distributed over the structure, we can take $\gamma_0$ outside the integral sign. Figure~\ref{fig:LDOS}b demonstrates the corresponding dependences for the cases in which there are emitting centers only in \textit{a}-Si$_{0.5}$C$_{0.5}$:H layers (solid curve in Fig.~\ref{fig:LDOS}b), or only in \textit{a}-SiO$_2$ layers (dash-and-dotted curve), and single-type emitting layers are uniformly distributed over the whole sample (dashed curve). It can be seen that two peaks appear in the spectrum at the frequencies corresponding to the outer boundaries of the PSB. However, the intensities of the peaks are strongly different, depending on which layers contain the emitting centers. For example, emission appears in the \textit{a}-Si$_{0.5}$C$_{0.5}$:H/\textit{a}-SiO$_2$  periodic structure under consideration only in \textit{a}-Si$_{0.5}$C$_{0.5}$:H layers, whereas, in fact, there is no emission from the silica layers. Accordingly, the integration should be only performed over the \textit{a}-Si$_{0.5}$C$_{0.5}$:H layers. As a result, the intensity of the peaks takes on a noticeable asymmetry. The long-wavelength peak is found to be substantially stronger than the short-wavelength peak. Figure~\ref{fig:ExpCalc}d presents the PL spectra calculated for the parameters of the structure, which correspond to the transmission spectra in Fig.~\ref{fig:ExpCalc}b.

\section{Discussion}

%%%%%%%%%%%%%%%%%%%%%%%%%%%%
%% Figure 7
\begin{figure}[!t]
\includegraphics{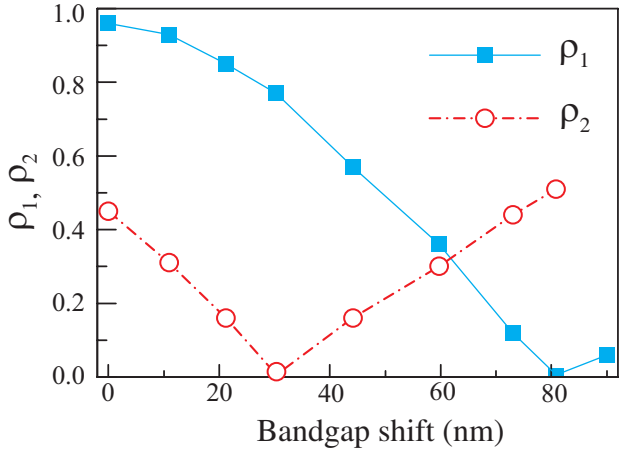}
\caption{Ratio of the difference of the PL intensities to the sum of these, measured at the wavelenght corresponding to the PSB edges calculated by formulas~(\ref{eq:diff_coef1},\ref{eq:diff_coef2}).
The cyan squares are values for the 1D PhC sample and red circles corresponds to the values for the \textit{a}-Si$_{0.5}$C$_{0.5}$:H reference film.
}
\label{fig:Diff}
\end{figure}
%
%%%%%%%%%%%%%%%%%%%%%%

It is important to take into account a number of circumstances in a study of how the radiative properties of emitters in PhC are modified. First, photonic crystals have a dielectric contrast associated with the distribution of several substances, each having its own radiative properties. To examine how the emission is modified for a particular kind of emitting centers, it is desirable that the PL would appear only in a single component of a PhC. In our study, we chose the technological regimes in which separate films were obtained and the PL excitation wavelength so that the \textit{a}-SiO$_2$ layers and the substrate yielded a negligible PL signal. Second, it is important that, on the one hand, the lifetime of a radiative center should strongly exceed the lifetime of a photon in the PhC. In this case, we remain in the framework of the weak-coupling approximation and the well-developed emission theory is applicable, as well as the description of the emission modification in terms of the Purcell effect. On the other hand, the PL line should be inhomogeneously broadened so as to cover the comparatively wide PSB in high-contrast PhCs. The hydrogenated amorphous alloy of silicon with carbon possesses the necessary properties. Third, a key factor in choosing the method for excitation of emitters within a PhC in PL studies in photonic-bandgap materials is their quantum efficiency \cite{vos2015localization}
\begin{equation}
\eta=\frac{\gamma_{rad}}{\gamma_{rad}+\gamma_{nrad}}
\end{equation}
where $\gamma_{rad}$ is the radiative recombination rate directly dependent of the photonic LDOS \cite{novotny2012principles}, and $\gamma_{nrad}$ is the nonradiative recombination rate. The quantum efficiency of emission from \textit{a}-Si$_{1-x}$C$_x$:H films with high content of carbon weakly depends on temperature and is several percent at $T=300 ^{\circ}$C, i.e., $\gamma_{rad}\ll1$\cite{siebert1987photoluminescence, carius2000optical}. In this case, $\gamma_{nrad}\gg\gamma_{rad}$ and the emission intensity is proportional to the  $\gamma_{rad}$-to-$\gamma_{nrad}$ ratio. Because $\gamma_{nrad}$ is determined only by the physicochemical properties of \textit{a}-Si$_{1-x}$C$_x$:H films, measurements of the spectral dependence of the PL intensity provides information about the LDOS under continuous-wave excitation conditions \cite{vos2015localization}. Thus, the choice of appropriate samples and PL excitation method made it possible to examine in the present study the effect of the LDOS on PL spectra.

Let us now consider the results of an LDOS simulation in terms of the 1D model. Comparison of Figs.~\ref{fig:ExpCalc}b and \ref{fig:ExpCalc}d shows that the calculated spectra reproduce the main features of the experimental curves. Consequently, the 1D model we used adequately reveals the nature of the observed effect, 
however we should keep in mind that our approach is restricted to analysis of features in the PL spectra detected in the normal direction of outgoing emission. We notice that the model does not account the nature of emission from a point source with a large portion of light escaping the structure in off-axis direction as well as coupling to guided-wave modes \cite{wubs2002local} and hence for calculating a directivity pattern of photoluminescence one has to perform computations using rigorous 3D methods \cite{fussell2004three, novotny1997allowed, wubs2002local}.  
However, to evaluate the measured specific intensity of emission in the on-axis direction it is sufficient to calculate only a part of Purcell factor and photonic LDOS that are exactly the quantities obtained from the 1D model.

Let us now consider Fig.~\ref{fig:LDOS} which shows how the 1D Purcell factor depends on wavelength and on the position of a point dipole source in the structure of the PhC. It can be seen that the maxima of the Purcell factor in \textit{a}-SiO$_2$ layers lie at the wavelength 600 nm at the outer boundary of the short-wavelength edge of the PSB ($\lambda=630$ nm), whereas the maxima in \textit{a}-Si$_{0.5}$C$_{0.5}$:H layers lie at wavelengths that are somewhat longer than the long-wavelength boundary of the PSB ($\lambda=810$ nm, Fig.~\ref{fig:LDOS}). The reason is that the eigenmodes in a PhC are so structured that the electric field is localized within layers with a higher dielectric index for bands at the low-frequency boundary of the PSB, and in layers with lower dielectric index at the high-frequency boundary \cite{g101}. The highest intensity maxima are observed far from the outer PhC boundaries, where the Bloch mode has enough space to be formed. It is noteworthy that the observed pattern is well consistent with the interference description of the Purcell effect for photonic modes \cite{rybin2016purcell}, according to which a large Purcell factor corresponds to the return (reflection) of the portion of wave out of phase with the source, and this leads to an additional decrease in energy in the oscillating source. A small Purcell factor corresponds to the reflection of the wave in phase with the source, with the result that energy is returned to the oscillator. On passing across the spectral range of the PSB, the phase of the Bloch wave undergoes a change by $\pi$. As a result, conditions are created at one of the PSB edges for a small Purcell factor to be formed, i.e., the reflected wave returns in phase, whereas at the other edge, the reflected wave returns out of phase (phase shift $\pi$) and the Purcell factor takes large values. 

Because only \textit{a}-Si$_{0.5}$C$_{0.5}$:H emits in our system, the PL spectra are dominated by the peak corresponding to the long-wavelength edge of the PSB. Nevertheless, the tails from the 1D Purcell factor maxima in \textit{a}-SiO$_2$ layers with small refractive index partly penetrate into the \textit{a}-Si$_{0.5}$C$_{0.5}$:H emitting layers. This results in that a short-wavelength peak appears, which has a lower intensity, but is noticeable against the background of the long-wavelength peak, and the more so against the frequencies in the PSB range (see Fig.~\ref{fig:LDOS}). In experimental studies, the PhC emission intensity is strongly affected by the frequency dependence of the emission rate of emitters in the \textit{a}-Si$_{0.5}$C$_{0.5}$:H material. To exclude this effect, we plotted against the spectral shift of the PSB center (Fig.~\ref{fig:Diff}) the dimensionless parameter $\rho_1$ defined as the ratio of the difference of the PL peak intensities at different PSB edges to the sum of these.
\begin{equation}
\rho_1=|I_{s1}-I_{l1}|/|I_{s1}+I_{l1}| \label{eq:diff_coef1}
\end{equation}
Here, $I_{s1}$ is the intensity of the PL peak corresponding to the short-wavelength edge of the PSB, and $I_{l1}$ is that of the PL peak corresponding to the long-wavelength edge of the PSB. In this case, the shift of the PSB was determined as the difference between the running and initial (665 nm) spectral positions of the stop-band center according to the plots in Fig.~\ref{fig:ExpCalc}a.

We also present the coefficient
\begin{equation}
\rho_2=\left| I_{s2}-I_{l2}\right|/\left| I_{s2}+I_{l2}\right| \label{eq:diff_coef2}
\end{equation}
for the PL spectrum of the reference \textit{a}-Si$_{0.5}$C$_{0.5}$:H film. In this case, the intensities $I_{s2}$ and $I_{l2}$ were taken from the PL contour of the film (red dash-and-dotted curve in Fig.~\ref{fig:ExpCalc}b) at the points corresponding to the spectral positions of the PL peaks in the \textit{a}-Si$_{0.5}$C$_{0.5}$:H/\textit{a}-SiO$_2$ multilayer structure. 
We notice that the shape of curves $\rho_1$ and $\rho_2$ are similar to each other. At the same time dependence resulted from emission of the 1D photonic crystal is significantly shifted to the longer wavelengths. This fact indicates the existence of additional effects that determine the shape and intensity of the luminescence spectrum. In this way, comparison
of the dependences obtained for $\rho_1$ and $\rho_2$ (Fig.~\ref{fig:Diff}) clearly confirms the conclusion that the variation of the specific intensity of the emission in the on-axis direction of PhC does not quantitatively follow the variation of the PL intensity from the single \textit{a}-Si$_{0.5}$C$_{0.5}$:H film, i.e., it is a consequence of the effect of the LDOS modification in the PhC.

\section{Conclusion}

In this study, we have fabricated by the PECVD technique 1D PhCs based on alternating quarter-wave \textit{a}-Si$_{0.5}$C$_{0.5}$:H/\textit{a}-SiO$_2$ layers, which can be regarded as model objects for studying the emission rate modification associated with the change in the photonic LDOS via engineering of the dielectric environment. Owing to the choice of materials and laser excitation wavelength, a structure was formed and studied, in which all the radiative centers are emitters of the same type (\textit{a}-Si$_{0.5}$C$_{0.5}$:H), which enabled an unambiguous interpretation of the experimental data obtained in the study. Thanks to the comparatively broad emission spectrum of \textit{a}-Si$_{0.5}$C$_{0.5}$:H, we could compare the modification of the photonic LDOSs at both PSB edges in a single experiment. The experimentally measured PL spectra have been described in terms of the 1D model and a good agreement was achieved between the experimental and calculated data. The theoretical model has revealed the nature of the asymmetry of the PL peaks at the PSB edges, which appears due to the sharp change in the 1D Purcell factor as a result of the phase jump of the Bloch wave by $\pi$ when the spectral range of the photonic stop-band is passed.

\begin{acknowledgments}
We thank Mikhail Voronov for fruitful discussions. M.R. acknowledges a support by the Ministry of Education and Science of the Russian Federation (Zadanie No. 3.1500.2017/Project part)
\end{acknowledgments}

%\bibliographystyle{revtex}
%\bibliography{aSiCH}

\end{document}